\documentclass[12pt]{iopart}
\usepackage{hyperref}
\usepackage{graphicx}
\usepackage{color}
\begin{document}

\title[Do moving clocks slow down?]{Do moving clocks slow down?}

\author{Abdaljalel Alizzi$^1$, Abhijit Sen$^1$ and Zurab K. Silagadze$^{1,2}$}
\address{$^1$ Novosibirsk State University, 630 090, Novosibirsk, Russia}
\address{$^2$  Budker Institute of Nuclear Physics, 630 090, Novosibirsk,
Russia}
\ead{abdaljalel90@gmail.com}
\ead{abhijit913@gmail.com}
\ead{silagadze@inp.nsk.su}

\begin{abstract}
The special theory of relativity has fundamentally changed our views of space and time. The relativity of simultaneity in particular, and the theory of relativity as a whole, still presents significant difficulty for beginners in the theory. The difficulty stems from the fact that the usual presentation of special relativity is based on Newtonian concepts, which are relativized and change their meaning in the course of the presentation. A better pedagogical practice, in our opinion, would be to base the presentation from the very beginning on the four-dimensional formulation of Minkowski and thus remove from the theory all the paradoxical connotations that invariably accompany the usual presentation of special relativity.

\end{abstract}

\section{Introduction}
In his groundbreaking paper \cite{Einstein:1905ve}, Einstein based his account of special relativity on two postulates, presumably assuming that these principles would be as robust and reliable as those of thermodynamics \cite{Brown}. Indeed, usually just this time-proven way of representing the special theory of relativity, with some variations, is used in many excellent textbooks (see, for example, \cite{SR1,SR2,SR3,SR4,SR5,SR6,SR7,SR8}). However, there are several reasons why such an approach cannot be considered a good pedagogy from the modern perspective (see \ref{appA}).

The real problem of the standard two-postulates approach and its varieties is that they view space and time as separate entities, while the real message of special relativity is that ``space by  itself  and  time  by  itself  will totally fade into shadows and only a kind of union of both will preserve independence" \cite{Minkowski}.

The three-dimensional standard approach to special relativity uses the decomposition of space-time into space and time based on inertial observers, hence the central role of Lorentz transformations in this approach. This limitation to the privileged class of inertial observers is too restrictive and unnecessary since the real world is made up of accelerated and rotating observers. Contrary to popular prejudice, there is no problem with treating such general observers in special relativity, albeit at the expense of a more complex mathematical formalism \cite{Gourgoulhon}. The relative complexity of the mathematical formalism is more than balanced by the fact that this approach greatly simplifies the study of general relativity. 

The root of the conceptual problems students have with the relativistic concept of time is that they confuse coordinate time with what they consider to be ``true time". Another difficulty is that they uncritically transfer the Newtonian concept of a reference frame into the theory of relativity. The great implication of general relativity is that coordinates play a secondary, albeit important (in practical calculations), role compared to the internal geometrical structure of space-time. Minkowski's four-dimensional formalism makes it possible to practice the same philosophy in the special theory of relativity.

In a four-dimensional picture, the only real, observer-independent time is the proper time. To introduce time coordinate beyond the observer's world line, a simultaneity convention is needed. We distinguish between the private time of a particular observer and the public time used and shared by a set of privileged observers. In inertial reference frames (more precisely, in inertial coordinate systems), privileged observers have parallel world lines and public time essentially coincides with private time. To illustrate that in the general case this is no longer true, we consider the Milne model, in which the world lines of inertial observers are not parallel, but intersect in one event (the Big Bang event).

Milne coordinates and their extension (Rindler coordinates) illustrate still another aspect of the four-dimensional formalism inherited from general relativity: the same space-time can be described by completely different coordinates, and it usually takes several different coordinate patches to cover the entire space-time (to label all events).

The importance of these considerations is in showing that by using only the concepts of proper time and Einstein simultaneity, it is possible to define inertial reference frame (as a congruence of parallel timelike worldlines filling all space-time), introduce coordinates (public time and public space), and we get the familiar presentation of special relativity. No need to use Einstein's postulates. The definition of public time and spatial coordinates with respect to inertial reference frame generates familiar formulas for relativistic interval and Lorentz transformations and in this case actually there is no need to differentiate between private and public times. However, if a different congruence is used to define privileged observers (as in the Milne model) and the corresponding public time they share, such a need arises.

The standard approach to teaching special relativity is similar to what Synge calls the ``cuckoo process" \cite{Synge}. First, students will learn that Newtonian concepts represent true physical reality. The relativistic eggs are then laid in a nest of Newtonian physics and processed in such a way that Newtonian concepts are suddenly mutated. Finally, the student is supposed to bite off his foster mother's head and fly out of the nest as a full-fledged relativist \cite {Synge}. 

However, this cuckoo process is not particularly effective: research shows that after completing regular training, students at all academic levels have significant difficulties with the relativity of simultaneity and the role of observers in inertial frames of reference, simultaneously believing in the ideas of absolute simultaneity and relativity of simultaneity that coexist harmoniously in their minds, not to mention other misconceptions \cite{miscon1,miscon2,miscon3}. Even among academic physicists, a poor understanding of special relativity is not uncommon \cite{Koks}.

In our opinion, better pedagogical practice would be to base the special relativity from the very beginning on the Minkowski four-dimensional formalism, which allows the introduction of relativistic concepts without any mention of their Newtonian counterparts \cite{Synge}. Then Newtonian concepts will emerge as perfectly sound and useful approximations under certain circumstances. Thus, we can avoid the situation faced with the traditional approach, then ``often the result is to destroy completely the confidence of the student in perfectly sound and useful concepts already acquired" \cite{bell_aspect_2004}.

One aspect of Minkowski's approach, the use of Minkowski diagrams has long been recognized as an effective teaching tool in special relativity \cite{Taylor,Shadowitz,Prado,Liu}, and no modern textbook on special relativity can avoid the use of four-vectors, another aspect Minkowski's approach. However, in this article, we are not interested in these things. Instead, we  will focus on only one feature of the four-dimensional formalism, namely how it clarifies the concept of time in the theory of relativity. We will try to demonstrate that the four-dimensional formalism, supplemented by the concepts of private and public times borrowed from Milne \cite{Milne}, makes the concept of time in special relativity crystal clear, leaving no room for a paradoxical aftertaste. 

The article uses natural units with the speed of light $c$ equal to one. If desired, $c$ can always be restored in formulas based on dimensional considerations.

\section{It is our turn to study special relativity}
The best motivating opening of the textbook, which we know, can be found in \cite{Goodstein}: ``Ludwig Boltzmann, who spent much of his life studying statistical mechanics, died in 1906, by his own hand. Paul Ehrenfest, carrying on his work, died similarly in 1933. Now it is our turn to study statistical mechanics. Perhaps it will be wise to approach the subject cautiously".

We are not aware of such a tragic fate associated with the study of the special theory of relativity. However, Herbert Dingle's sad story is dramatic enough to be remembered in this context \cite{Chang,Kevin}. For our purposes, the Dingle case is of twofold interest:
\begin{enumerate}
    \item It demonstrates that the traditional presentation of the theory of relativity can confuse even professional scientists.
    \item It demonstrates problems in communication (understanding each other's arguments) when discussing scientific topics, even in such a logical and rigorous discipline as physics.
\end{enumerate}
Therefore, in \ref{AppB} we provide a  rather detailed description of the Dingle affair. In the end, the main result of this controversy was the complete destruction of Dingle's scientific reputation. The most heartbreaking aspect of Dingle's case is that, towards the end of his life, he disowned almost all of his friends and associates, whom he began to regard as accomplices in the worldwide fraud in favor of the theory of relativity, after he failed to convince them in their moral and ethical responsibility to join him in his crusade against special relativity \cite{Kevin}.

Indeed, it seems it will be a wise advice for students of special relativity to approach this subject with caution.

\section{Time in special relativity: proper, private and public}
Physics as a scientific discipline is organized to consider various systems such as elastic media, electromagnetism, weak and strong interactions, gravity, ideal fluids, condensed matter systems, etc. In spite of numerous differences in details, these systems have some common features. Namely, in order to describe them, various fields are introduced that live in a space-time manifold, the dynamics of these fields is described by partial differential equations, which require an initial-value formulation for predicting the behavior of the fields \cite{Geroch:1996kg}.

All fundamental partial differential equations in physics are of hyperbolic type, elliptic and parabolic systems arising as mere approximations of hyperbolic systems. More precisely, the vast majority of equations describing physical systems can be represented as a system of symmetric hyperbolic partial differential equations, and each such system deﬁnes its own causal cones with respect to which an initial value formulation determines the domain of causal influence \cite{Geroch:1996kg,Geroch:2010da}.

Gravity, as described by Einstein equations, can also be represented by a system of symmetric hyperbolic partial differential equations and hence has its own causal cones \cite{Geroch:1996kg,Geroch:2010da}. Special relativity, as a limiting case of general relativity, inherits these causal cones, which in this case are null cones of the Minkowski metric. 

Our current knowledge of the laws of nature is consistent with the notion that the physical world is organized around a commonality of causal cones, with the cones of special relativity playing a preferable role: other causal cones either coincide with the cones of special relativity, like the light cones of electromagnetism, or lie within them, like the causal cones of various condensed matter systems \cite{Geroch:2010da}. Although we cannot rule out the existence of hidden sectors whose causal cones lie outside the null cones of special relativity \cite{Geroch:2010da,Chashchina:2021jrx}, the special role of Minkowski null cones (commonly called light cones because they coincide) for known physical systems is obvious.  Therefore, it is desirable to emphasize from the very beginning this special role of light cones in the teaching of special relativity, instead of emphasizing the role of Einstein's second postulate.

The standard presentation of special relativity overemphasizes the relative, coordinate-dependent aspects of the theory (relativity of simultaneity, time dilation, length contraction, Lorentz transformations, etc.). However, the real physical meaning has only coordinate-independent quantities. As Felix Klein noted in 1910, ``What modern physicists call the theory of relativity is the invariant theory of the four-dimensional space-time domain $x,y,z,t$ (the Minkowskian universe) with respect to a particular group of collineations -- namely, the Lorentz group" \cite{Klein:1910bea}. An approach to special relativity that emphasizes from the outset the role of absolute rather than relative concepts was initiated by Alfred Robb \cite{Robb4,Robb1,Robb2,Robb3}.

The absolute concepts {\it before} and {\it after} are defined as follows. For any event $A$ in space-time, the events lying in the past light cone $A$ occur before the event $A$, and the events lying in the future light cone $A$ occur after $A$. However, there are events that occur neither before nor after $A$ (events outside the light cone with apex at $A$). In Minkowski's terminology, they occur {\it elsewhere}, but it would be more correct to say that those events simply cannot be causally related to $A$. 

Therefore, the light-cone structure defines a partial conical order in space-time \cite{Robb4}. To make this order quantitative, the concept of an ideal clock is introduced, which measures proper time along its time-like world line (the "length" of this world line).

Proper time plays the same central role in special relativity as Newtonian absolute time in classical physics, and in this sense it is its successor, but a very peculiar successor. Newtonian mechanics can be formulated in four-dimensional form, much like special and general relativity \cite{Havas:1964zza}. However, the classic proper times are combined together to create the global master clock, and ``the physics of matter listens to the ticks of this global clock. In relativity, by contrast, one can often define a global time (see below), but matter-energy doesn't listen to it. Restricted to the clocks to which the physics listens, the ideal master clock of classical physics has shattered into a huge plurality of miniature ideal clocks in relativity, one for every time-like path" \cite{Callender}.

However, the proper time is determined by an ideal clock only along clock's world line. For a given particle (observer) with an ideal clock, to extend the concept of time  beyond its world line, we need the concept of {\it simultaneity}. Nature does not provide an absolute concept of simultaneity, unlike the concepts of {\it before}, {\it after} and {\it elsewhere}. Simultaneity must be defined by a convention (by a stipulation, as Einstein called it \cite{Einstein:1905ve}).

The definition of Einstein's simultaneity is as follows. Suppose the world line of an observer is some time-like curve $\gamma$. He sends a light pulse at his proper time $\tau_1$ to some nearby event $A$. At $A$, the light pulse is immediately reflected back and received by the observer at the proper time $\tau_2$. According to Einstein, an event $B$ on $\gamma$ is simultaneous with $A$ if it corresponds to the proper time $\tau=\frac{1}{2}(\tau_1+\tau_2)$. We can also define the distance between simultaneous events $A$ and $B$ as $l=\frac{1}{2}(\tau_2-\tau_1)$ \cite{Robb4}. 

Since $\tau=\frac{1}{2}(\tau_1+\tau_2)$, $l^2=(\tau_2-\tau)(\tau-\tau_1)$. This form, if extended even for non-simultaneous events when $\tau\ne\frac{1}{2}(\tau_1+\tau_2)$, makes it possible to introduce the concept of ``distance" (relativistic interval) to all sufficiently close events  $A$ and $B$ \cite{Geroch1,Geroch2,Synge,Salecker:1957be}:
\begin{equation}
s^2=(\tau_2-\tau)(\tau-\tau_1).
\label{eq1}
\end{equation}
Einstein simultaneity allows us to ascribe time to events that are in the immediate vicinity of some referential event. Namely, let the referential event $O$ lie on the world line $\gamma$ of the observer $S$ and it corresponds to the proper time $\tau=0$ on $\gamma$. The radar coordinates $(t,x)$ (in two-dimensional space-time) of an event $A$ in an infinitesimal neighborhood of $O$ are defined as follows \cite{Robb4,Synge1921}:
\begin{equation}
t=\frac{1}{2}(\tau_1+\tau_2),\;\;\;\;x=\frac{1}{2}(\tau_2-\tau_1),
\label{eq2}
\end{equation}
where the meanings of $\tau_1$ and $\tau_2$ were described above. Therefore, the tangent to the world line $\gamma$ at the point $O$ determines the natural time-direction for $S$, and the time coordinate $t$ is equal to the proper time $\tau$ of the event $B$ on $\gamma$, which is simultaneous with $A$, while the coordinate $x$ is the distance from $A$ to $B$.

The specified procedure with light signals used in these definitions is generally local in the sense that in the general space-times it works only for sufficiently close events \cite{Perlick:2007np}. However, for inertial observers with straight world lines in Minkowski (flat) space-time, the radar coordinates are well defined in the whole space-time and determine the foliation of Minkowski space-time into private time $t$ and the private space of the observer $S$. 

In special relativity, the inertial frame of reference is, in fact, a set of freely floating ideal clocks with parallel world lines filling the space-time \cite{Schild}. These observers (clocks) share with $S$ the same foliation of Minkowski space-time. This circumstance allows us to elevate the private time and private space of the observer $S$  to the status of public time and public space of the inertial frame (the congruence of world lines parallel to $\gamma$), which we denote also by $S$.

\begin{figure}[ht]
\begin{center}
\includegraphics[width=0.7\textwidth]{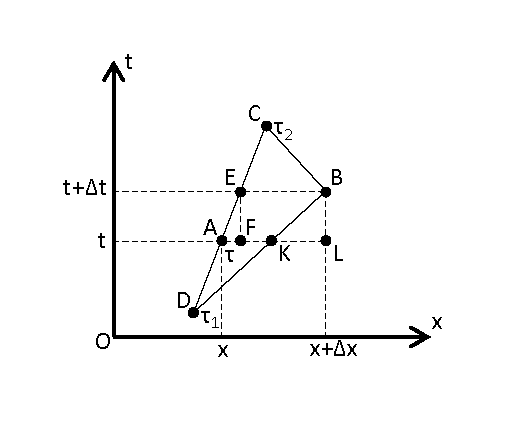}
\end{center}
\caption{An illustration of the calculation of the relativistic interval between infinitesimally close events.}
\label{fig1}
\end{figure}
Only proper times are used in Robb-Geroch's definition of relativistic interval (\ref{eq1}). Requiring observer-independence of this definition, it can be shown that the standard Minkowski geometry does follow, which becomes evident by expressing the relativistic interval in public coordinates of some inertial reference frame. As was shown in Ref. \cite{Sagaydak:2021qth}, the definition of the relativistic interval (\ref{eq1}) can be generalized in such a way that it will lead not to Minkowski geometry, but to Finsler geometry. We will not consider this more general possibility here, but will use the ideas of \cite{Sagaydak:2021qth} when considering the technical details.

In the reference frame $S$, let us express the squared interval $\Delta s^2$ between two infinitesimally close events $A$ and $B$ with coordinates $(t,x)$ and $(t+\Delta t, x+\Delta x)$ through coordinate differences $\Delta t$ and $\Delta x$. To do this, we choose some world line $DAEC$ passing through the event $A$ (see Fig.\ref{fig1}), which can be considered a straight line segment, since we are considering infinitely close events. If the world line $DAEC$ is characterized by the velocity $\beta$ in coordinates $t,x$ and if it makes an angle $\theta$ with the $x$ axis, then
\begin{equation}
\tan{\theta}=\frac{1}{\beta},\;\;\;\sin{\theta}=\frac{1}{\sqrt{1+\beta^2}},\;\;\;\cos{\theta}=
\frac{\beta}{\sqrt{1+\beta^2}}.
\label{eq3}
\end{equation}
Suppose a photon is emitted at proper time $\tau_1$ (event $D$) and after reflection at event $B$ is received at proper $\tau_2$ (event $C$). We also assume that event $A$ corresponds to proper time $\tau$ on the world line $DAEC$. Since $DB$ is the world line of the photon and, thus, tilted at an angle of $\pi/4$, $KL=BL= \Delta t $. Besides, $\angle KDA=\theta-\pi/4$, $\angle AKD=\pi/4$, and from the triangle $ADK$, $AD/\sin{\frac{\pi}{4}}=AK/\sin{\left (\theta-\frac{\pi}{4}\right )}$. Therefore,   
\begin{equation}
AD=AK\,\frac{\sin{\frac{\pi}{4}}}{\sin{\left (\theta-\frac{\pi}{4}\right )}}=\frac{\Delta x-\Delta t}{\sin{\theta}-\cos{\theta}}=\frac{\sqrt{1+\beta^2}}{1-\beta}\,\left (\Delta x-\Delta t\right ).
\label{eq4}
\end{equation}
A coefficient $k(\beta)$, which converts the Euclidean length of the $AK$ segment 
into the corresponding proper time, may depend on the world line and, therefore, depends on the velocity that characterizes this world line in the inertial reference frame $S$. Then from equation (\ref{eq4}) it follows that
\begin{equation}
\tau-\tau_1=k(\beta)\,AD=k(\beta)\,\frac{\sqrt{1+\beta^2}}{1-\beta}\,\left (\Delta x-\Delta t\right ),   
\label{eq5}
\end{equation}
Guided again by Fig.\ref{fig1}, we can calculate $\tau_2-\tau$ in the similar way. Since $\angle CEB =\theta$ and $\angle EBC =\pi/4$, the theorem of sines in the triangle $ECB$ gives $EC/\sin{\frac{\pi}{4}}=EB/\sin{\left(\pi-\theta-\frac{\pi}{4}\right )}$. On the other hand, $EF=\Delta t$, $AF=EF\,\cot{\theta}=\beta\Delta t$ and $EB=\Delta x-AF=\Delta x-\beta\Delta t$. Therefore, 
\begin{equation}
\hspace*{-5mm}
EC=\left (\Delta x-\beta\Delta t\right ) \frac { \sin { \frac{\pi} {4}}}{\sin{\left (\theta+\frac{\pi}{4}\right )}}=\frac{\Delta x-\beta\Delta t}{\cos{\theta}+\sin{\theta}}=\frac{\sqrt{1+\beta^2}}{1+\beta}\,\left (\Delta x-\beta\Delta t\right ).
\label{6}
\end{equation}
But $AC=AE+EC=EF/\sin{\theta}+EC$. Therefore,
\begin{equation}
AC=\sqrt{1+\beta^2}\,\Delta t+\frac{\sqrt{1+\beta^2}}{1+\beta}\,\left (\Delta x-\beta\Delta t\right )=\frac{\sqrt{1+\beta^2}}{1+\beta}\,\left (\Delta x+\Delta t\right ),
\label{eq7}
\end{equation}
and
\begin{equation}
\tau_2-\tau=k(\beta)\,AC=k(\beta)\,\frac{\sqrt{1+\beta^2}}{1+\beta}\,\left (\Delta x+\Delta t\right ).
\label{eq8}
\end{equation}
Combining equations (\ref{eq1}), (\ref{eq5}) and (\ref{eq8}), we get
\begin{equation}
\Delta s^2=k^2(\beta)\,\frac{1+\beta^2}{1-\beta^2}\,(\Delta x^2-\Delta t^2).
\label{eq9}
\end{equation}
The $\Delta s^2$ interval between the events $A$ and $B$ must not depend on the world line $DAEC$ used in its calculation. This will be so if $\beta$-dependent factor in (\ref{eq9}) is actually a constant, fixed to be one by the requirement $k(0)=1$. Then
\begin{equation}
k(\beta)=\sqrt{\frac{1-\beta^2}{1+\beta^2}},
\label{eq10}
\end{equation}
and the interval (\ref{eq9}) takes its conventional form $\Delta s^2=\Delta x^2-\Delta t^2$.

In inertial reference frames, the public time and public space coincide to the private time and private space of a particular observer from the congruence that constitutes this frame of reference. In the general case, this is not true, since public time and public space depend on the congruence of the world lines of observers who share these concepts, and they can differ from the private time and private space of a particular observer from this congruence, as the Milne model clearly demonstrates. 

\section{Milne model and Rindler coordinates} 
The Milne model describes a quarter of the same Minkowski space-time in different coordinates. The same is true for the Rindler coordinates, which describe another part of the Minkowski space-time. These coordinates differ from ordinary inertial coordinates because a different set of privileged observers are used to define public time and public space. Details of the Milne model can be found in \cite{Milne,Chashchina:2014gsa}. Here we use it just to demonstrate the difference between private and public times (and private and public spaces).

\begin{figure}[ht]
\begin{center}
\includegraphics[width=0.5\textwidth]{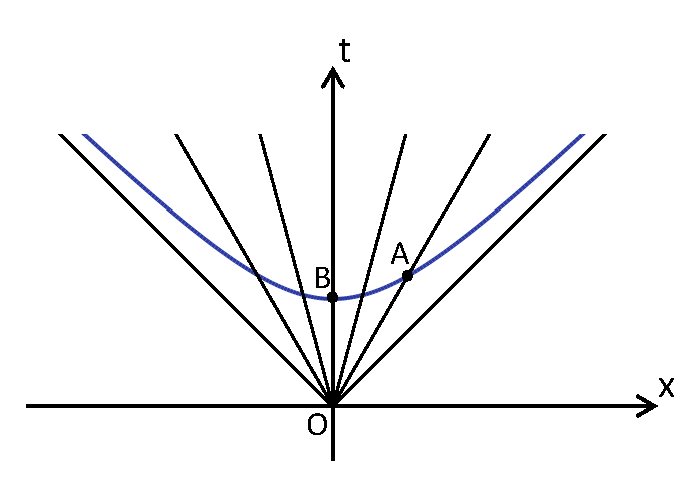}
\end{center}
\caption{Illustration of the Milne model.}
\label{fig2}
\end{figure}

Fundamental particles (observers) of the Milne model are all inertial observers whose world lines intersect at one point (the event $O$, see Fig.\ref{fig2}) in the Minkowski space-time. The following interpretation is possible: during the $O$ event (Big Bang event), an explosion occurs, which ejects an infinite number of fundamental observers in radial directions with the entire range of constant velocities from zero to one (the speed of light).

The Milne space-time is a part of the Minkowski space-time (one quarter of it) obtained by reducing the Poincaré symmetry of the Minkowski space-time to its subgroup (the Lorentz group), which leaves the Big Bang event $O$ invariant.  This peculiar symmetry of the congruence of the world lines of fundamental observers dictates a convenient choice of public time: we can define remote simultaneity in this case not by Einstein's convention, but by requiring the same proper times for all fundamental observers, provided that the Big Bang corresponds to zero proper time.

Since the fundamental observers move in Minkowski's spatiotemporal background, for each observer we can still introduce a private time and a private space that permeate Minkowski's entire world. However, when limited to the Milne world, private time and private spatial coordinates cover only a finite interval: private time is not negative, and for a given private time $t$, private spatial coordinates can take values only from $-t$ to $t $. In other words, Milne's fundamental observers occupy a finite ball in private space, whose radius increases in private time at the speed of light.  

It can be shown that for a very special kind of velocity distribution of fundamental observers, the Milne universe will satisfy the cosmological principle, that is, it will look the same for all fundamental observers \cite{Milne,Chashchina:2014gsa}.

Let's choose one fundamental observer as ``at rest" and associate with him private coordinates with origin at the point of the Big Bang. The public time coordinate (cosmic time) of the event $A$ with private coordinates $(t,x)$ is equal to $T=\tau=t\sqrt{1-\beta^2}=\sqrt{t^2-x^2}$, where $\beta=x/t$ is the velocity of the fundamental observer $A$ in the particular private coordinate system chosen by us. Then the axis $T=\mathrm{const}$ of the public space is the hyperbola $t^2-x^2=\tau^2$, and the public spatial coordinate $X$ of the event $A$ is the Lorentzian length of the segment $AB$ of this hyperbola (see Fig.\ref{fig2}. The event $B$ lies on the $t$-axis and therefore has private coordinates $(\tau,0)$). Along the hyperbola, $dt/dx=x/t=x/\sqrt{\tau^2+x^2}$. Therefore, 
\begin{equation}
X=\int\limits_B^A\sqrt{dy^2-dt^2}=\int\limits_0^x \frac{\tau}{\sqrt{\tau^2+y^2}}\,dy=\tau\,\mathrm{arsinh}\,{\frac{x}{\tau}}. 
\label{eq11}
\end{equation}
Then $x=\tau\sinh{(X/\tau)}$ and $t=\sqrt{\tau^2+x^2}=\tau\cosh{(X/\tau)}$. These relations show that instead of $X$ the dimensionless quantity $\psi=X/\tau$ can be used as a spatial coordinate. In this case, Minkowski coordinates $(t,x)$ and Milne coordinates $(T,\psi)$ are related as follows:
\begin{equation}
t=T\cosh{\psi},\;\;\; x=T\sinh{\psi},\;\;\;\beta=\frac{x}{t}=\tanh{\psi}.
\label{eq12}
\end{equation}
The last equation shows that in Milne space-time the natural dimensionless spatial coordinate $\psi$ represents the fundamental observer's rapidity from the Minkowski space-time point of view (the rapidity, defined by $\beta=\tanh{\psi}$, is the natural Lorentz group parameter \cite{Levy_Leblond_1981}).

In the case of a three-dimensional space, $x$ in (\ref{eq12}) is replaced by the radial coordinate $r$ and for the square of the interval, introducing the usual spherical angles $\theta$ and $\phi$, we obtain \cite{Chashchina:2014gsa}
\begin{equation}
\hspace*{-22mm}
ds^2=dr^2+r^2(d\theta^2+\sin^2{\theta}\,d\phi^2)-dt^2=T^2\,[d\psi^2+\sinh^2{\psi}\,(d\theta^2+\sin^2{\theta}\,d\phi^2)]-dT^2,
\label{eq13}
\end{equation}
which indicates that the Milne universe is mathematically equivalent to an expanding Friedmann-Lema\^{i}tre-Robertson-Walker (FLRW) universe with vanishing total energy density and vanishing cosmological constant \cite{Milne,Chashchina:2014gsa}.

Milne's simultaneity hyperboloids implement Lobachevsky (hyperbolic) spatial geometry \cite{Reynolds}, demonstrating the relativity of space in the most dramatic way: a spatial slice of the same Minkowski space-time has a Euclidean geometry with an ordinary foliation, and a negatively curved hyperbolic geometry with a Milne foliation.

However, the Milne foliation (coordinates) covers only one quarter of Minkowski space-time, which can be expanded to the opposite quarter also by simply reversing the sign of the cosmic time coordinate $T$ (which replaces the expanding Milne universe with the contracting Milne universe and the Big Bang singularity with the Big Crunch singularity). What about the other half of Minkowski space-time?

The straight lines in this other half that pass through the Big Bang event are spacelike, not timelike, and therefore cannot serve as observers' world lines. However, they can serve as a multitude of spatial axes for some congruence of observers. In fact, this situation is very peculiar, since it implies that the Big Bang event occurs simultaneously with each event on the world line of each such (Rindler) observer. These observers cannot be inertial, because all spatial axes of the inertial observer are parallel to each other and therefore cannot intersect. 

\begin{figure}[ht]
\begin{center}
\includegraphics[width=0.5\textwidth]{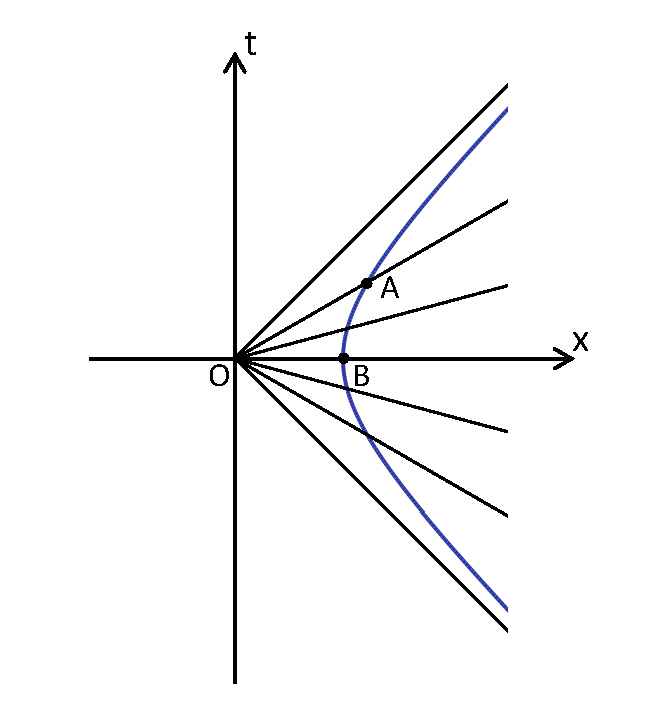}
\end{center}
\caption{Illustration of the Rindler right edge.}
\label{fig2a}
\end{figure}

Let the event $A$ has coordinates $(t,x)$ in the inertial reference frame of the observer, which we assumed to be at rest when working with the Milne universe. If the world line of the Rindler observer passes through $A$ and is described by the functional relation $t=t(x)$, then the vector $(t,x)$ directed along the spatial axis of this observer must be Lorentz orthogonal to the tangent vector of the world line at point $A$, which is parallel to the vector $\left(\frac{dt}{dx}\,dx,dx\right)$. This condition gives $t\frac{dt}{dx}=x$ and, therefore, the world line is described by the relation $x^2-t^2=X_R^2$, where $X_R$ is some constant (see Fig.\ref{fig2a}). Actually, $1/X_R$ is the proper acceleration of this observer \cite{Rindler:1966zz}.

As the time coordinate $T_R$ of the event $A$, we can use the proper time corresponding to the segment $AB$  (see Fig.\ref{fig2a}):
\begin{equation}
T_R=\int\limits_B^A \sqrt{-ds^2}=\int\limits_B^A \sqrt{d\tau^2-dx^2}=\int\limits_0^t\frac{X_R}{\sqrt{X_R^2+\tau^2}}\,d\tau=X_R\,\mathrm{arsinh}\,{\frac{t}{X_R}}.
\label{eq14}
\end{equation}
Therefore, $t=X_R\sinh{(T_R/X_R)}$ and $x=\sqrt{t^2+X_R^2}=X_R\cosh{(T_R/X_R)}$. Then $\beta=dx/dt=\tanh{(T_R/X_R)}$. This shows that $\psi=T_R/X_R$ is rapidity and it can be used as a convenient time coordinate in Rindler space-time instead of $T_R$. The spatial coordinate of the event $A$ in Rindler space-time is $X_R$. Indeed, since $\beta=\tanh{\psi}=t/x$, we get $k(\beta)\,OA=\sqrt{(1-\beta^2)/(1+\beta^2)}\,\sqrt{t^2+x^2}=\sqrt{x^2-t^2}=X_R$.
As we see,  Minkowski coordinates $(t,x)$ and Rindler coordinates $(\psi,X_R)$ are related as follows:
\begin{equation}
t=X_R\sinh{\psi},\;\;\; x=X_R\cosh{\psi}.
\label{eq15}
\end{equation}
Compared to the Milne case, the roles of space and time are reversed, which is very reminiscent of Kruskal's famous extension of the Schwarzschild space-time \cite{Rindler:1966zz} (the relation between the Milne and Rindler coordinates can also be found in \cite{Eriksern_2004}, where a very interesting discussion of the electromagnetic field of a uniformly accelerated charge is presented from the point of view of the Rindler and Milne observers).

It is clear from (\ref{eq15}) that Rindler coordinates cover only $x>0$, $|t|<x$ portion of the Minkowski space-time (right Rindler wedge). Reversing sign of $X_R$, we get left Rindler wedge. Such division into Rindler wedges arises only in the two-dimensional space-time. As in the Milne case, when space is three-dimensional, $x$ in (\ref{eq15}) is replaced by the radial coordinate and we can define spherical Rindler coordinates $(t_R,r_R,\theta,\phi)$ as follows 
\begin{eqnarray} &&
t=r_R\sinh{\psi},\;\;\;x=r_R\cosh{\psi}\sin{\theta}\cos{\phi},\nonumber \\ &&
y=r_R\cosh{\psi}\sin{\theta}\sin{\phi},\;\;\;z=r_R\cosh{\psi}\cos{\theta}.
\label{eq16}
\end{eqnarray}
In these coordinates, squared interval takes the form \cite{Chashchina:2014gsa}
\begin{equation}
ds^2=dr_R^2+r_R^2\cosh^2{\psi}\left (d\theta^2+\sin^2{\theta}\,d\phi^2\right )-r_R^2\,d\psi^2,
\label{eq17}
\end{equation}
and they cover the whole external part of the Big Bang’s light-cone.

In summary, it is a characteristic property of Milne's observers that all of their time axes pass through one event (the Big Bang), while, on the other hand, a characteristic property of Rindler's fundamental observers is that their spatial axes all pass through this event. Together, the Milne and Rindler coordinates cover the entire Minkowski space-time. 

The Milne model illustrates some of the implications of general relativity that, unfortunately, are rarely discussed in the context of special relativity. Namely, coordinates are simply labels of space-time points (events) and do not always have a spatiotemporal meaning. In case the coordinate chart covers only a part of space-time, we need another charts (an atlas of coordinate charts) to work with the entire space-time, and these charts can differ significantly from each other.

Inertial coordinates are just a very convenient choice in Minkowski space-time, as they are adapted to the symmetries of that space-time and encompass all of space-time. However, we can use any other coordinates, for example, definition of simultaneity other than Einstein's. In this case, all coordinate-dependent quantities, such as the one-way speed of light, relativistic formulas for addition of velocities, length contraction and time dilation, Lorentz transformations between two inertial observers, relativistic expressions for energy and momentum, can undergo radical changes, so that phrases that are usually considered inherent features of special relativity, such as {\it moving clocks go slow} and {\it simultaneity is relative}, will turn out to be wrong \cite{Leubner_1992}. All coordinate-independent quantities characterizing genuine physical effects, such as differential aging of twins, remain, of course, indifferent to such a change in coordinization.

In Milne coordinates, every fundamental observer is at rest, but three-dimensional space is negatively curved and expanding. On the contrary, in inertial coordinates, three-dimensional space is the usual static Euclidean space, but the fundamental observers move in a certain characteristic way.

Can this replacement of space expansion by certain motion patterns of fundamental observers be achieved in other, more realistic FLRW cosmologies as well? Infeld and Schild investigated this question in a remarkable paper \cite{Infeld:1945mio}. They showed that for each FLRW cosmology there is a so-called cosmological coordinate system, an analogue of inertial public coordinates, in which the FLRW metric is conformally equivalent to the Minkowski metric. Such coordinates allow to discard all discussions of curved, expanding universes and shift the responsibility for cosmological phenomena essentially to the patterns of motion of fundamental observers (or, if we insist on a geometric rather than kinematic interpretation, to the gauge field of Weyl's integrable geometry, which determines the behavior of clocks and measuring rods) \cite{Infeld:1945mio}.

In FLRW coordinates, the speed of light in any fixed direction depends on cosmic time and the geometry of a constant curvature three-dimensional space. In cosmological coordinates, the speed of light is always and in any direction equal to one (with our choice of units), as in Minkowski space-time.

\section{Lorentz transformations}
Lorentz transformations relate the coordinates $(t,x)$ of the event $A$ in the inertial frame $S$ with the coordinates $(t^\prime,x^\prime)$ of the same event in the inertial frame $S^ \prime$. To obtain them, consider the observer $S^\prime$, which moves with the velocity $\beta$ relative to $S$ and passes the common referential event $O$ at $t=0$. The new time direction $t^\prime$ in the reference frame $S^\prime$ coincides with the direction of the world line of this observer. Let us prove that the spatial axis $x^\prime$ is symmetric to $t^\prime$ with respect to the diagonal $t=x$, and thus the Einstein simultaneity convention guarantees that the speed of light in all inertial reference frames is equal to one.
\begin{figure}[ht]
\begin{center}
\includegraphics[width=0.5\textwidth]{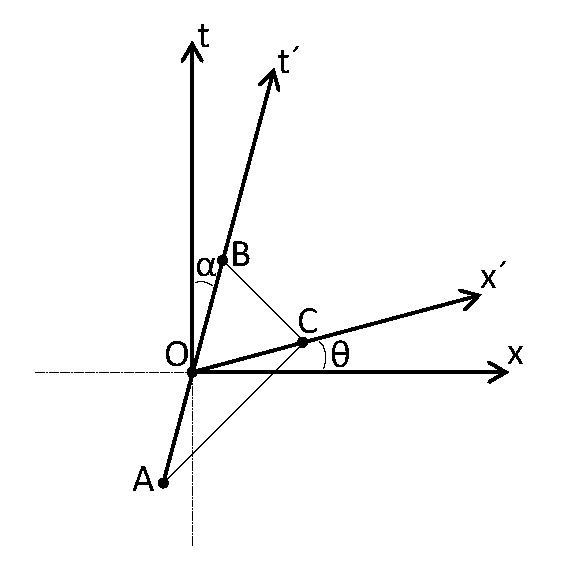}
\end{center}
\caption{Construction of the $x^\prime$ axis.}
\label{fig3}
\end{figure}
If the event $C$ is on the $x^\prime$ axis, then events $O$ and $C$ are simultaneous and thus $AO=OB$, where $AC$ and $CB$ are photon world lines (see Fig.\ref{fig3}). Since the photon world lines are inclined at an angle $\pi/4$ with respect to the $t$ axis, then $\angle BCA=\pi/2$, $\angle OAC=\pi/4-\alpha$ and $\angle OBC =\pi/4+\alpha$ (to get the last two equations, imagine vertical lines parallel to the $t$ axis at $A$ and $B$). The theorem of sines in the triangle $AOC$ gives $AO/\sin{\left(\frac{\pi}{2}-\gamma\right)}=OC/\sin{\left(\frac{\pi}{4}-\alpha\right)}$, and in the triangle $OCB$ $OB/\sin{\gamma}=OC/\sin{\left(\frac{\pi}{4}+\alpha\right)}$, where $\gamma=\angle BCO$. Since $AO=OB$ we get
\begin{equation}
\frac{\sin{\gamma}}{\sin{\left(\frac{\pi}{2}-\gamma\right)}}=\frac{\sin{\left(\frac{\pi}{4}+\alpha\right)}}{\sin{\left(\frac{\pi}{4}-\alpha\right)}}=\frac{\sin{\left[\frac{\pi}{2}-\left(\frac{\pi}{4}-\alpha\right)\right]}}{\sin{\left(\frac{\pi}{4}-\alpha\right)}}=\frac{\cos{\left(\frac{\pi}{4}-\alpha\right)}}{\sin{\left(\frac{\pi}{4}-\alpha\right)}}.
\label{new_eq}
\end{equation}
Therefore, $\tan{\gamma}=\cot{\left(\frac{\pi}{4}-\alpha\right)}$ and $\gamma=\frac{\pi}{2}-\left(\frac{\pi}{4}-\alpha\right)=\left(\frac{\pi}{4}+\alpha\right)$. Then $\angle BOC=\pi-\left(\frac{\pi}{4}+\alpha\right)-\left(\frac{\pi}{4}+\alpha\right)=\frac{\pi}{2}-2\alpha$ and $\theta=\frac{\pi}{2}-\alpha-\left(\frac{\pi}{2}-2\alpha\right)=\alpha$, which ends the proof.

\begin{figure}[ht]
\begin{center}
\includegraphics[width=0.5\textwidth]{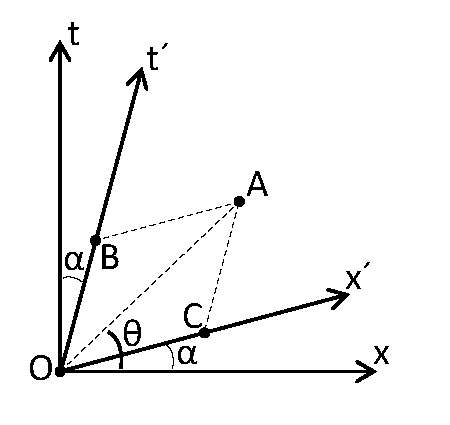}
\end{center}
\caption{Figure explaining how the Lorentz transformations are obtained.}
\label{fig4}
\end{figure}

The coefficient that converts the Euclidean length along the $x^\prime$ axis to the corresponding coordinate coincides with the coefficient that converts the Euclidean length along the $t^\prime$ axis to the corresponding proper time, since in radar coordinates the spatial distance is defined in terms of proper time. Therefore, $t^\prime=k(\beta)\,OB$, $x^\prime=k(\beta)\,OC$, while $t=OA\sin{\theta}$, $x=OA\cos{\theta}$, where $\theta$ is the inclination angle of $OA$ relative to the $x$ axis (see Fig.\ref{fig4}). 

\begin{figure}[ht]
\begin{center}
\includegraphics[width=0.7\textwidth]{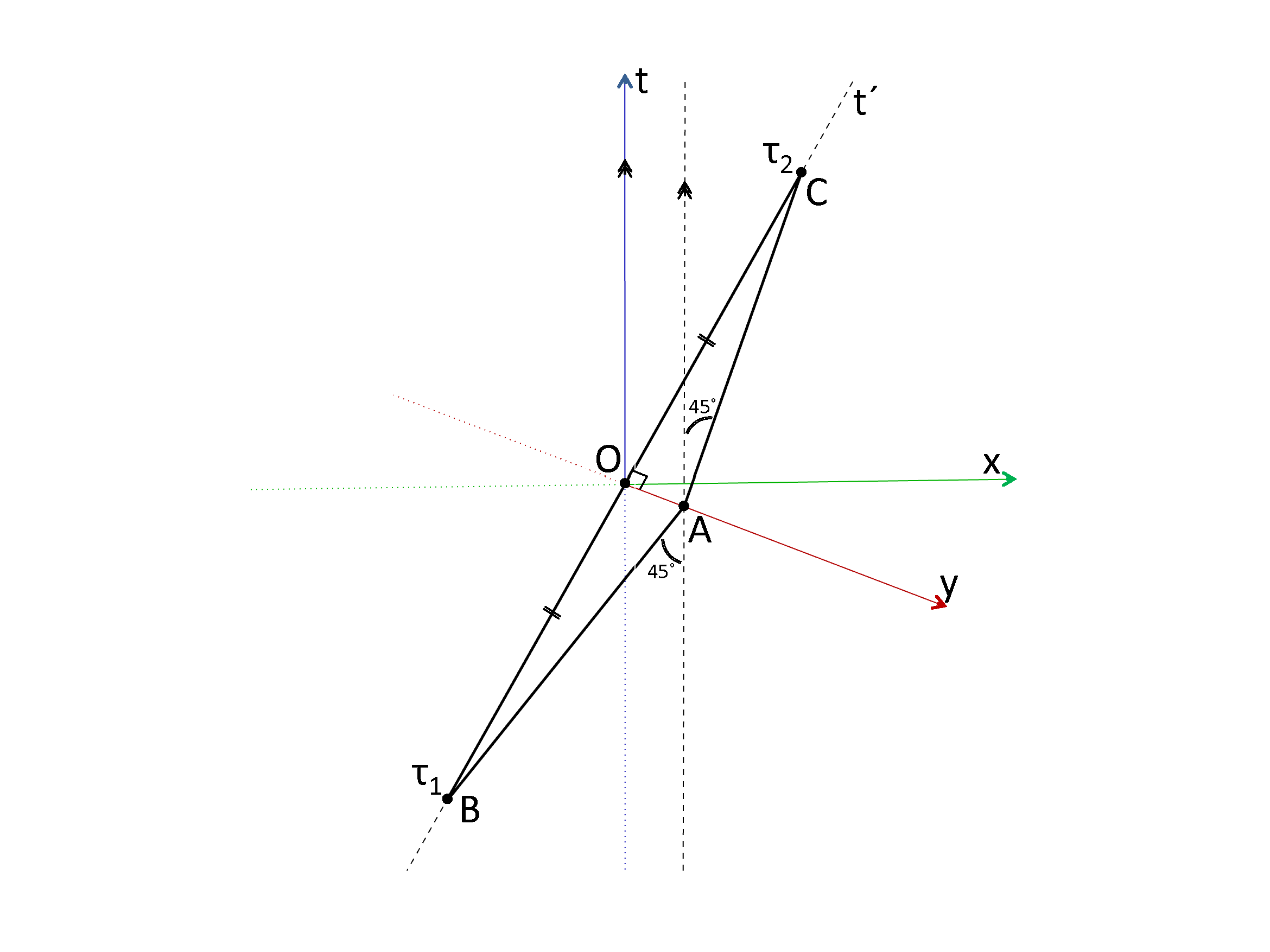}
\end{center}
\caption{Figure explaining how transverse coordinate transformations are obtained.}
\label{fig5}
\end{figure}

It is clear from the Fig.\ref{fig4} that $\angle BOA=\angle OAC=\frac{\pi}{2}-(\theta+\alpha)$, $\angle AOC=\theta-\alpha$, and $\angle OCA=\pi-\angle OAC-\angle AOC=\frac{\pi}{2}+2\alpha$.  In the triangle $OAC$, $OA/\sin{\left(\frac{\pi}{2}+2\alpha\right)}=OC/\sin{\left(\frac{\pi}{2}-(\theta+\alpha)\right)}$.
Therefore, using $\tan{\alpha}=\beta$ and hence $\sin{\alpha}=\beta/\sqrt{1+\beta^2}$, $\cos{\alpha}=1/\sqrt{1+\beta^2}$, we get
\begin{equation}
\hspace*{-10mm}
OC=OA\,\frac{\cos{(\theta+\alpha)}}{\cos{2\alpha}}=OA\,\frac{\cos{\theta}\cos{\alpha}-\sin{\theta}\sin{\alpha}}{\cos^2{\alpha}-\sin^2{\alpha}}=\frac{\sqrt{1+\beta^2}}{1-\beta^2}\,(x-\beta t),
\label{eq18}
\end{equation}
and
\begin{equation}
x^\prime=k(\beta)\,OC=\frac{x-\beta t}{\sqrt{1-\beta^2}}=\gamma\,(x-\beta t),\;\;\;
\gamma=\frac{1}{\sqrt{1-\beta^2}}.
\label{eq19}
\end{equation}
where we have used (\ref{eq10}). Similarly, in the triangle $OBA$, $OA/\sin{\angle OBA}=OB/\sin{\angle BAO}$. However, $\angle OBA=\angle OCA=\frac{\pi}{2}+2\alpha$, $\angle BAO=\angle AOC=\theta-\alpha$, and
hence
\begin{equation}
\hspace*{-15mm}
OB=OA\,\frac{\sin{(\theta-\alpha)}}{\sin{\left (\frac{\pi}{2}+2\alpha\right )}}=OA\,\frac{\sin{\theta}\cos{\alpha}-\cos{\theta}\sin{\alpha}}{\cos^2{\alpha}-\sin^2{\alpha}}=\frac{\sqrt{1+\beta^2}}{1-\beta^2}\,(t-\beta x).
\label{eq20}
\end{equation}
Therefore,
\begin{equation}
t^\prime=k(\beta)\,OB=\gamma\,(t-\beta x).
\label{eq21}
\end{equation}
We can find the transformation laws of the transverse coordinates too. To do this, let us consider an event $A$ with coordinates $(0,0,y,0)$ in the $S$ reference frame that lies on the $y$-axis. The events $A$ and $O$ are simultaneous for all observers  moving along the $x$ axis, since $t_A=x_A=0$. Therefore, for the observer $S^\prime$ with the world line $BOC$ (see Fig.\ref{fig5}) $BO=OC$, since $\tau_1=-k(\beta)\,BO$, $\tau_2=k(\beta)\,OC$ and
$t^\prime_A=\frac{1}{2}(\tau_1+\tau_2)=0$, with $\beta$ being the velocity of the observer $S^\prime$ in the $S$ reference frame.

Since $AC$ is the photon world line, the Euclidean vector $\overrightarrow{AC}$ with coordinates $(t_C,\beta t_C,-y,0)$ must make the angle $\pi/4$ with time axis $t$.  Therefore,
\begin{equation}
\frac{\overrightarrow{AC}\cdot(1,0,0,0)}{|\overrightarrow{AC}|}=\frac{t_C}{\sqrt{t_C^2(1+
\beta^2)+y^2}}=\cos{\frac{\pi}{4}}=\frac{1}{\sqrt{2}}.
\label{eq22}
\end{equation}
The solution with respect to $t_c$ is $t_C=y/\sqrt{1-\beta^2}$, with the second solution $t_B=-y/\sqrt{1-\beta^2}$ corresponding to the event $B$. Therefore, 
\begin{equation}
OC=\sqrt{t_C^2(1+\beta^2)}=\sqrt{\frac{1+\beta^2}{1-\beta^2}}\,y, 
\label{eq23A}
\end{equation}
and
\begin{equation}
\tau_2=k(\beta)\,OC=y.
\label{eq23}
\end{equation}
Coordinate $y^\prime$ in the reference frame $S^\prime$ is the distance between the simultaneous events $A$ and $O$ that equals to $y^\prime=\frac{1}{2}(\tau_2-\tau_1)=\tau_2$. Then (\ref{eq23}) indicates that $y^\prime=y$.

As we see, the Lorentz transformations have their usual form 
\begin{equation}
t^\prime=\gamma\,(t-\beta x), \;\;\;
  x^\prime=\gamma\,(x-\beta t), \;\;\;
 y^\prime=y, \;\;\; z^\prime=z.
\label{eq24}
\end{equation}

Our consideration of the Lorentz transformations and the relativistic interval closely follows \cite{Sagaydak:2021qth}, where a more general situation was considered (Finslerian generalization of the Minkowski metric). We reproduce all the details here only to make the present paper self-contained, so that  ``the reader will not need to have his fingers at eleven places to follow an argument" \cite{Veltman}.

\section{Conclusions}
Minkowski's four-dimensional formulation removes completely a paradoxical flavour from the special relativistic definition of time. ``Proper time is the most physically significant, most physical real time we know. It corresponds to the ticking of physical clocks and measures the natural rhythms of actual events" \cite{Misner}. Other times in relativity are derivative, conditional concepts. 

The four-velocity of an observer defines a natural time-direction for that observer. Using Einstein's simultaneity and this direction of time, the observer splits space-time into private time and private space in his immediate vicinity. In Minkowski space-time and in some highly symmetric space-times, private time-and-space coordinates can be extended to the entire space-time, thus providing public time-and-space coordinates. 

The usual interpretation of the results of the Hafele-Keating experiment and many measurements of the mean lifetime of muons in the atmosphere is that a moving clock ticks slower than a stationary clock,  and the student wonders, like Dingle, what is happening to the moving clock. Minkowski's four-dimensional perspective clearly shows that the ``slowdown" is not due to some physical mechanism acting on the moving clock. All ideal clocks in Minkowski or any other general relativistic space-time are identical regardless of their state of motion (regardless of the timelike paths they follow in space-time), they simply measure the proper time along the clock's world line. This is the definition of an ideal clock, and just as a ruler does not change when we measure the length of a curve in Euclidean space, an ideal clock does not change in any way when we measure proper time along a curved time-like world line.

It is important that students firmly grasp the distinction between proper time (which is a frame-independent quantity) and coordinate time (frame-dependent public time shared by a certain class of privileged observers). Ideal clocks indicate the proper time along their world lines, and in relativity there is no objective way to compare their ticking rates unless the world lines merge into a finite segment (unless the clocks are in the same place and mutually stationary for some duration). Coordinate time is subjective as it requires conventions about simultaneity, choice of space-time coordinate system, etc. to define it. The rate of a clock with respect to the coordinate time of some space–time reference system, of course, depends on the chosen frame of reference, and this is what is usually meant when one speaks of slowing down of a moving clock. It is true, however, that to carry out real calculations, it is necessary to make a choice of coordinates, and a convenient choice of coordinates can greatly simplify calculations \cite{Fromholz:2013hka}. 

Many conceptual confusions emerge because students do not think ``geometrically" in terms of the intrinsic structure of space-time, but rather in terms of coordinates. It took even Einstein quite some time to appreciate the elegance and power of Minkowski's geometric approach, in which coordinates lose their privileged role and even their immediate physical meaning \cite{Giovanelli_2021}.

To summarise, in our opinion, the best pedagogical practice for teaching special relativity with the postulated Lorentz symmetry would be to rely entirely on Min\-kowski's four-dimensional formalism and avoid at all such phrases as ``moving clocks slow down", which only confuse students. It will be even better if some key ideas of general relativity are introduced first, and special relativity is treated as a special case \cite{Rindler:1994,Hartle:2005cq}. To paraphrase L\'{e}vy Leblond, in the past hundred years, we have accumulated sufficient familiarity, theoretical as well as experimental, with the relativistic world to no longer look at it through Newtonian glasses \cite{Levy_Leblond_1981}. Some additional concluding remarks can be found in \ref{AppC}.

\section*{Acknowledgments}
We thank anonymous referees for constructive comments.  The work is supported by the Ministry of Education and Science of the Russian
Federation.

\appendix
\section{Drawbacks of standard approach to special relativity}
\label{appA}
First, and most importantly, from the outset, it lends a paradoxical flavor to our most  
fundamental and well-established theory, which, combined with quantum mechanics, forms the 
basis of modern physics. This is due to the fact that Einstein's postulates are formulated in Newtonian terms, such as the inertial frame of reference and the one-way speed of light (which involves measuring space and time intervals with a meter and a clock), and in Newtonian reading, the two postulates clearly contradict each other. 

Modern versions of the second postulate are themselves paradoxical. ``To take as a postulate that the speed  of light is constant relative  to  changes  in  reference frame is to assume an  apparent absurdity.  It  goes  against common  sense.  No  wonder,  thinks  a  student,  that  we  can derive  other  absurdities,  such  as  time  dilation and length contraction, from  the  premises" \cite{Baierlein}. 

In fact, Einstein's formulation of the second postulate, unlike modern formulations, does not imply that the speed of light is the same in all inertial reference frames. Instead, Einstein declares that the speed of light does not depend on the speed of the source, which was a completely natural and in no way paradoxical statement for contemporaries who considered light as a wave in the ether. The exact same phrase was used by Wilhelm Wien even earlier, in 1904 \cite{Baierlein}. It is the combination of the two postulates that seems impossible to Newtonian eyes. 

Note that throughout his life, Einstein was increasingly concerned about how he phrased his 1905 paper, and he emphasized in his 1949 autobiographical notes that ``the universal principle of the special theory of relativity is contained in the postulate: The laws of physics are invariant with respect to the Lorentz-transformations" \cite{Brown:2009vca}.

The standard presentation of the special theory of relativity emphasizes the special role of light (electromagnetic radiation). This is understandable since Maxwell's classical electrodynamics was our first relativistically invariant theory. However, we now know that Lorentz invariance is valid for all natural phenomena and that light propagates at an invariant velocity because the photon is massless. There is no compelling theoretical requirement for the photon to be strictly massless \cite{Okun:1991nr}, although  experimentally photon mass is very tightly constrained \cite{Wang:2021nrl}. The masslessness of the photon stems from a special pattern of electroweak symmetry breaking.  Nonetheless, the tiny mass of the photon would neither destroy the perfect agreement of the slightly modified Standard Model and experiment \cite{Okun:1991nr}, nor would it undermine the validity of special relativity, but that would be enough to deny the photon a special role in relativity. In Einstein's own words, "the  Lorentz  transformation,  the real basis of the special relativity theory, in itself has nothing to do with  the  Maxwell theory" \cite{bams}.  

Both of these shortcomings (the paradoxical nature of the premises and the dependence of relativity on the properties of light) are eliminated in von Ignatowsky's approach, in which the invariant velocity is not postulated, but derived from less paradoxical and favorable premises, such as the homogeneity of space and time and the isotropy of space, as well as the relativity principle \cite{Ignatowsky,Frank:1911omv,Torretti,Berzi:1969bv} (further references can be found in \cite{Baccetti:2011aa}).

In von Ignatowsky's approach, like the standard presentation of special relativity, measuring rods and clocks are used to give operational meaning to coordinates in the inertial frame, and if we allow measuring rods to change their length when gently put in a uniform motion, the whole theory follows from the reciprocity principle along with some ``self evident" assumptions \cite{Silagadze:2007eb}.

One may wonder whether it is legitimate to use measuring rods as the basis of the special theory of relativity, since a direct consequence of this very theory is the impossibility of having absolutely rigid bodies. However, ``the entities introduced in the course of the logical exposition of a theory are theoretical, abstract, ideal, conceptual entities; the question of their relation to the actual, physical, concrete, real entities whose behavior the theory is intended to help understand is a  complex one"  \cite{Stachel}. The legitimacy of the introduction of measuring rods as basic elements can be based on the concept of rigid inertial motion, which is not prohibited by special relativity under appropriate conditions \cite{Stachel}.

\section{The Dingle Affair}
\label{AppB}
Herbert Dingle was not an ordinary anti-relativistic crackpot, but a distinguished English philosopher of science and physicist. However, his path to science was not easy \cite{Kevin, MacTutor}. 

Dingle was raised by his mother, as his father died shortly after his birth. He had to drop out of school at the age of 14 and worked as a clerk for the next 11 years. Despite the difficult economic situation of the family, Dingle continued to educate himself and at the age of 25 received a scholarship at the prestigious Imperial College London. 

After graduating from college, Dingle specialized in spectroscopic astronomy. He was elected a Fellow of the Royal Astronomical Society in 1922, became a professor, and then head of the spectroscopy section at Imperial College. Later he served as President of the Royal Astronomical Society from 1951 to 1953. 

Dingle had a strong interest in the philosophy of science, which soon began to prevail over spectroscopy. He was one of the founders of the British Society for the History of Science, which he chaired from 1955 to 1957. In 1948 he founded the Philosophy of Science Group, which later became the British Society for the Philosophy of Science. Dingle was also the founder of the British Journal of the Philosophy of Science, and vice president of the International Union for the History of Science.

Dingle was introduced to the theory of relativity by renowned British philosopher and mathematician Alfred North Whitehead, co-author with Bertrand Russell of the epoch making {\it Principia Mathematica}, well-known for his formidable achievements in the foundations of mathematics and logic. This may have been an unfortunate choice, since Whitehead was first and foremost a philosopher and secondly a mathematical physicist with his own unorthodox views of space-time and relativity.  Whitehead even created his own relativistic theory of gravity, which for a long time competed with Einstein's general theory of relativity \cite{Synge:2005yh,Gibbons:2006jy}.

As a result, as a retrospective reading of Dingle's writings on special relativity indicates, he had always fundamentally misunderstood relativity, even such basic concepts as the definitions of an event and an inertial coordinate system, as well as the meaning of the relativity of simultaneity \cite{Kevin}. The latter is perhaps unsurprising, if we remember that Dingle was introduced to the theory of relativity by Whitehead, who wrote in 1922 ``I maintain the old-fashioned belief in the fundamental character of simultaneity. But I adapt it to the novel outlook by the qualification that the meaning of simultaneity may be different in different individual experiences" \cite{Whitehead}.

Despite his unorthodox views, Dingle was generally considered an expert in special relativity. In 1922 he wrote a popular book ``Relativity for All" and in 1940 a monograph “The Special Theory of Relativity”. In 1932 he visited the California Institute of Technology as a Rockefeller Foundation Scholar, there he met Einstein and worked on relativistic cosmology with Richard Tolman, an author of well-known book  ``Relativity, Thermodynamics and Cosmology" \cite{Kevin,Chang}. 

It seems that from the beginning Dingle mixed up the theory of relativity with Leibniz's relationalism \cite{Kevin}. He writes in his 1940 monograph ``The Principle of Relativity may be stated thus: There is no meaning in absolute motion" \cite{Dingle}. This, of course, is not the principle of relativity, according to which the laws of physical phenomena should be the same only for a special class of inertial observers. 

Dingle's 1940 book on the theory of relativity was criticized by Epstein \cite{Epstein}, and for good reason: Dingle in this book denies the reality of time dilation and bases the entire theory of special relativity solely on length contraction as he understands it (also not canonically). In the book, Dingle reiterates his previous argument that the special theory of relativity gives no information, and makes no assumption, about the behavior of moving clocks \cite{Dingle_Nature}.

The argument is the following. In an hourglass, the elapsed time can be measured with equal accuracy by the number of fallen grains, by their total mass, or by their total volume. Since the number of grains, mass, and volume have different transformation laws for moving frames of reference, it is unclear how a moving hourglass behaves, since the first measurement method will not show retardation, while others will show slowing down, but different from each other. 

Every successful modern student of relativity should be able to understand that Dingle's argument is flawed. Since it means that if two hourglasses are nearby, one measuring time by counting the number of fallen grains, and the other measuring time by measuring the mass of the fallen sand, show the same time at each moment for an observer at rest relative to the hourglasses, they will show different readings for a moving observer. But ``this is clearly not true: if the distance between the clocks is negligible: if the two dials look the same, at any given instant, to one observer, they look the same to all, no matter what mechanism is behind them" \cite{Newman}.

The important thing to realize is that you do not compare the rates of a moving clock and a given fixed clock. What is actually done is that the readings of the moving clock are compared with the readings of {\it two} synchronized fixed clocks, and this is where the relativity of simultaneity comes into play and causes time dilation for all three versions of Dingle's hourglass. 

Unfortunately, Dingle's opponents were not always clear enough about the intended meaning of their rebuttal. For example, according to Epstein \cite{Epstein} Dingle's clocks, which measure time by the mass or volume of the collected sand, are illegitimate clocks. While according to Infeld \cite{Infeld} they, of course, ought to be called clocks, but because these clocks have no rhythm (are aperiodic in Infeld's terminology), the question whether or not they show changing rhythm when in uniform motion is meaningless. 

Unsurprisingly, Dingle remained unconvinced (however, we do not understand why he did not appreciate Newman's very clear and concise argument \cite{Newman}). An exchange of views followed in the form of journal articles \cite{Chang,Craig}, but it only ended in confusion \cite{Chang}. 

Despite its misleading nature and criticism from the mainstream, Dingle's book on special relativity was a success among readers, as evidenced by its four editions. This success is probably due not to the book's scientific merit, but to Dingle's eloquent literary style, which was the result of his keen interest in English literature. For example, Dingle describes Minkowski's four-dimensional approach as follows: ``The Minkowski four-dimensional continuum is essentially metaphorical. When we represent a change in our standard of rest by a rotation of axes, we are leaving prose for poetry. We are doing what Shakespeare did when he represented the illumination of mountain peaks by the morning sunlight as jocund day standing tiptoe on the misty mountain tops. The representation is beautiful, and scientifically permissible if we do not forget that it is symbolic. We may extend it with profit if we bear this in mind, but at great peril if we do not" \cite{Dingle}.

Dingle's real problem with relativity began in 1956, when he embarked on a long crusade, first against the generally accepted interpretation of the twin paradox, and then against special relativity itself. 

For a classical description of the twin paradox, see, for example, \cite{Schild,Marder} (the literature on the twin paradox is huge, we will additionally point out only some articles \cite{Fremlin_1980,Eriksen_1990,Low_1990,Soni_2002,Pesic_2003,Gron_2006,Boblest_2011}, published at various times in this journal). The paradox dates back to Einstein's first paper on relativity \cite{Einstein:1905ve}. Einstein considered two clocks, not twins. In his brilliant 1911 lecture in Bologna, Paul Langevin began his talk by asking if any of them would like to ``dedicate two years of his life to find out what Earth would look like in two hundred years".  All one had to do for this, continued Langevin, was to travel at a speed close to the speed of light \cite{Canales+2015}. Weyl was probably the first who mentioned the twins in this context \cite{Reichenberger:2018}. However, neither Einstein, nor Langevine, nor Weyl called this special relativistic effect a paradox. It acquired a paradoxical connotation only after the philosophical works of Bergson \cite{During,Canales+2015}, who attended Langevin's lecture in Bologna, became interested in the philosophical aspects of the special relativistic concept of time, found Einstein's definition of time in terms of clocks completely aberrant \cite{Canales+2015} and got into a heated debate with Einstein about this after they met in Paris in 1922 \cite{During,Canales+2015}.

Initially, Dingle, like Bergson, did not question the validity of special relativity. But they believed that time dilation was only an apparent effect, caused simply by a convention on how to define distant simultaneity. According to Bergson, ``in the Theory of Relativity the slowing-down of clocks by their displacement is, rightfully, {\it as real} as the shrinkage of objects in terms of distance" \cite{Canales+2015}.  There is no ambiguity in simultaneity, then the twins unite again. So, Dingle argued, time dilation,  as a purely conditional effect, does not appear, and therefore the real prediction of the theory of relativity, according to Dingle, is that there will be no age difference between the twins.  

In fact, we can find rationalizations for Dingle's arguments that are worth discussing as they can improve our understanding of special relativity. Dingle's objection to the conventional interpretation of the twin paradox boils down to three arguments \cite{Chang}: 
\begin{itemize}
\item  Symmetry argument: two twins are symmetrical; The acceleration of the moving twin has nothing to do with the intended time dilation effect.
\item Conventionality Argument: time dilation is a conventional effect that appears only when the concept of distant simultaneity is involved.
\item Absurdity Argument: the interpretation of the behavior of the Earth-bound twin by the accelerated twin is physically impossible.
\end{itemize}
Dingle is correct that acceleration does not play an essential role in the twin paradox. Lord Halsbury, Dingle's only significant ally in the twin paradox debate, made this quite clear when he proposed a no-acceleration version of the paradox in 1957. In fact, this version was suggested much earlier (in 1927) by Luise Lange  and looks as follows: ``to deal with motions free throughout from accelerations we avoid the jump from one system to the other; instead, we leave Paul attached to his first system and place another observer, James, on the one moving oppositely, who sets his clock in accordance with Paul's when he passes him" \cite{Lange}. When James meets Peter, Paul's Earth-bound twin, and compares his watch to Peter's, the same asymmetrical aging effect is found.

Asymmetric aging still persists in compact spaces where both twins can be inertial observers, but nevertheless have the ability to compare their ages more than once when their paths cross \cite{Barrow:2001rj,Uzan_2002}. Again, acceleration plays no role in breaking the symmetry between twins.

Herman Bondi's apt analogy explains well why Dingle is both right in neglecting the role of acceleration in the twin paradox problem and wrong in assuming that the twins are in a symmetrical situation. Bondi indicates that  the twin paradox situation is similar to the situation with two motorists who traveled from a common starting point to the same destination, but on different routes. ``Suppose our second motorist follows a route consisting of a number of straight segments joined by short, sharp corners. Of course his route will be longer than the route of the motorist driving along the straight road. His route will be longer {\it because} it curves, but the actual length of the curves themselves is quite negligible compared with the extra length of his drive. In other words, and this is the essential point of the problem, although the extra length is {\it due} to the curves it does not lie {\it in} the curves“ \cite{Chang}.

From a space-time perspective, the twin paradox is trivial and not paradoxical in any sense. The crucial difference between the twins is their different trajectories in space-time. In Euclidean geometry, a straight line corresponds to the shortest distance between two points, while in Minkowski geometry, a straight line corresponds to the longest proper time. In the canonical version of the paradox, acceleration serves simply as a marker for a world line that is not straight and therefore corresponds to less overall aging. Considering the twin paradox in compact spaces, it is more difficult and subtle to find such a marker \cite{Uzan_2002,Roukema:2006yu}.

Dingle is also right that time dilation is largely a conventional effect. During inertial motion, each twin is right in saying that the other’s clock runs slow, ``which would be impossible if the effect were something that happened to a clock instead of to a judgment of simultaneity" \cite{Chang}. What is not conventional is the Minkowski geometry of space-time, revealed for inertial observers through the Lorentz transformations that lead to the specific form of time dilation equations. This is why time dilation is not a completely conventional effect and can lead to absolute effects like asymmetric aging of twins (the first clock effect).

An ideal clock that measures proper time is the most primitive, fundamental concept in relativity. In fact, it plays the role of the unit of ``length" (interval) along a timelike world line in Minkowski geometry and, like a unit of length in Euclidean geometry, which does not change when measuring the length of the Euclidean curve, the ideal clock does not slow down or speed up when we measure the proper time of a timelike curve in Minkowski space-time. In both special and general relativity, speaking about measuring proper times and proper lengths, it is more or less tacitly assumed that, ``to put it somewhat provocatively, there exist clocks that do not slow down and rods that do not shorten in any circumstances" \cite{Giovanelly}.

We can compare rates of two ideal clocks only when they are located side by side and are mutually motionless. Any comparison of rates of remote ideal clocks or clocks in relative motion necessarily involves the concept of distant simultaneity and thus includes a conventional element. What is experimentally verified when comparing the rates of ideal clocks is the absence of the second clock effect: if two identical coincident clocks at relative rest tick the same way, they will tick the same way even after they reunite again after passing through different time-like world lines in space-time \cite{Lobo:2018zrz}.

An experimental method for testing whether a given clock is an ideal clock (that is the clock measuring proper time) using only light rays and freely falling particles is described in \cite{Perlick:1987}.

A curved space-time analogue of the first clock effect is the so called gravitational time dilation,  which recently attracted attention as a pedagogical tool for explaining free fall \cite{Gould_2016,Czarnecka:2020kfh,Stannard_2016}. 

An ideal clock in curved space-time also measures proper time and never slows down or speeds up. When in the special or general theory of relativity they talk about time dilation, they mean that the standard clock is compared with coordinate time, which is conventional and less fundamental than proper time. It is not always recognized that gravitational time dilation is a particularly delicate topic \cite{Scott_2015}: it is assumed (and experimentally verified for transverse accelerations of about $10^{18}g$ \cite{Bailey:1977de}) that acceleration does not affect the rate of an ideal clock (the clock hypothesis), then how can gravity affect the rate in light of Einstein's principle of equivalence?

In fact, an essential aspect of both gravitational time dilation and special relativistic time dilation is the path dependence of the proper time, experimentally tested in the famous experiment of Hafele and Keating \cite{Hafele}.

As for the absurdity argument, Builder correctly noted \cite{Builder:1957nj} that the strange behavior of the coordinates of the Earth-bound twin from the point of view of the moving twin during the deceleration-acceleration phase of the latter is not a problem, since these strange features are simply an illustration of the artificiality of the coordinate system  of the accelerated observer, which cannot be realized physically. An exchange of views took place between Dingle and Builder, in which Dingle did not comment on the absurdity argument at all \cite{Chang}.

In special theory (or in general theory) of relativity, there is no such thing as {\it a coordinate system of an accelerated observer}.  In their renowned textbook on general relativity, Misner, Thorne and Wheeler noted: ``It is very easy to put together the words {\it the coordinate system of an accelerated observer}, but it is much harder to find a concept these words might refer to. The most useful first remark one can make about these words is that, if taken seriously, they are self-contradictory" \cite{Misner}.

However, one can find fairly good coordinates that cover a part of space-time, which are relevant when discussing the paradox of twins, and in this sense replace the coordinate system of the accelerated observer. The use of such coordinates is necessary for traveler twin to calculate the elapsed proper time of the twin bound to the Earth, since two different inertial coordinate systems associated with the twin moving outward and towards the Earth do not cover the entire world line of the terrestrial twin between encounters \cite{Marder}. For example, one can use Kottler-Rindler coordinates \cite{Gamboa}, or M\"{a}rzke-Wheeler coordinates \cite{Pauri:2000cr}. 

In his review of Dingle's last anti-relativistic book ``Science at the Crossroads", Ian Roxburgh, a British astronomer and mathematician, writes: ``The surprising aspect of the Dingle affair is that he has some sort of a hearing" \cite{Roxburgh}.
 
Indeed, science as practiced by scientists is not at all like to the image that an enthusiastic freshman student might have in mind, that science deals only with the truth, which is revealed by guesswork and refutation. In reality "there is a powerful establishment and a belief system. There are power seekers and career men, and if someone challenges the establishment he should not expect a sympathetic hearing" \cite{Roxburgh}.

Yes, Dingle had some sort of a hearing, thanks to his past academic achievements and regalia. At first, the exchange of views between Dingle and his opponents was intense, and from what was said above, one could reasonably expect that the controversy should have ended with some kind of enlightenment, since many very intelligent people were involved in it. However, enlightenment, if any, was very limited. As E.~G.~Cullwick remarked in 1959, ``On one thing Professor Dingle’s critics are all agreed, that he is wrong. They do not all agree, however, on the nature of his error" \cite{Chang}. As analysis of the exchanges shows, after more than a year and a half of intensive, detailed and seemingly thorough exchanges, there was still an incredibly low level of understanding of each other's arguments \cite{Chang}. Not surprisingly, this misunderstanding led ``to otherwise clear-headed researchers talking past one another", as Craig Callender aptly noted on another occasion \cite{Callender}.

\section{Additional concluding remarks}
\label{AppC}
In his 1922 Kyoto lecture ``How I Created the Theory of Relativity," Einstein describes the decisive moment when he became enlightened in conversations with his friend Michel Besso: ``My interpretation was really about the concept of time. Namely, time could not be defined absolutely, but is in an inseparable relationship with the signal velocity" \cite{Kyoto_Lecture}. 

Although the alleged cause of Einstein’s redefinition of time is not known for certain, and  it is conceivable that Einstein, consciously or unconsciously, may have borrowed from other authors, most plausibly from Poincar\'{e}, more than his writings and sayings suggest \cite{Darrigol_2021}, without any doubt, this new definition of time was the most shocking and paradoxical aspect of special relativity. 

Einstein elevated Lorentz's local time to the status of ``true" time, and for his contemporaries it became the successor to Newtonian time. This immediately gave the theory a paradoxical tinge, since Newtonian time is absolute, while Lorentz's local time varies depending on the inertial frame of reference.

When the traditional exposition of special relativity, following the early Einstein, elevates public (coordinate) time to the status of ``true" time, it misses the mark and opens a Pandora's box of paradoxes in the fledgling minds of students educated in Newtonian physics. If it is said that moving clocks run slowly, the student, like Dingle, would invariably want to know  ``what would slow them down, physically" \cite{Chang}.

In fact, Dingle was quite right when he claimed that ``nothing at all happens to the moving clock" \cite{Chang}. Good clocks always measure the proper time regardless of their state of motion, unless their acceleration or tidal forces are so great as to interfere with their inner workings. 

The real question is why we can even find real clocks (such as atomic clocks) that work very much like the ideal clocks of relativity. The short answer to this question is that a real clock behaves like an ideal clock because the physics that governs its inner workings is Lorentz invariant.  And the long answer is never ending scientific exploration, in which Dingle's question about the origin of Lorentz invariance is a legitimate question for which we still do not have an answer. 

Dingle's question about what slows down a moving clock does not make sense in Einstein's version of relativity because, firstly, ideal clocks do not ``slow down" in the sense that Dingle suggests, and secondly, Lorentz invariance is postulated, not derived from a more fundamental theory. 

However, this question makes sense in the more ambitious program of Lorentz and Poincar\'{e} on how the theory of relativity should be developed. Lorentz noted in 1909 that the main difference between the two programs is that ``Einstein simply postulates what we have deduced, with some difficulty and not altogether satisfactorily, from the fundamental equations of the electromagnetic  field" \cite{Zahar}. 

Although the Lorentz-Poincar\'{e} program of deriving the Lorentz symmetry rather than postulating it was never completely abandoned \cite{Janossi,Brown}, it is clear that Einstein's program replaced it \cite{Zahar,Zahar1}, and for good reason: even today, Lorentzian symmetry cannot be deduced from a more fundamental theory. 

The phrase ``moving clocks slow down" is universally used and has been in common use for over a hundred years, and we realize that our suggestion not to use it, while correct, is not very realistic due to the "intellectual inertia" \cite {Wilczek_2004} of the physics community. This intellectual inertia is well confirmed by Okun's unsuccessful struggle with outdated concept of relativistic mass \cite{Okun,Silagadze:2011yj}. Nevertheless, we decided to write this article, following an advice from the speech of Walter Kotschnig delivered at Smith College on November 8, 1939:

``Let us keep our minds open by all means, as long as that means keeping our sense of perspective and seeking an understanding of the forces which mould the world. But don’t keep your minds so open that your brains fall out! There are still things in this world which are true and things which are false; acts which are right and acts which are wrong, even if there are statesmen who hide their designs under the cloak of high-sounding phrases".

\section*{References}
\bibliography{SR_time}

\end{document}